# Modification of the Magnetic Properties of Co$_2$Y Hexaferrites by Divalent and Trivalent Metal Substitutions

Sami H. Mahmood[1,a*], Muna D. Zaqsaw[1,b] Osama E. Mohsen[1,c], Ahmad Awadallah[1,d], Ibrahim Bsoul[2,e], Mufeed Awawdeh[3,f], Qassem I. Mohaidat[3,g]

[1]Physics Department, The University of Jordan, Amman 11942, Jordan
[2]Physics Department, Al al-Bayt University, Mafraq 13040, Jordan
[3]Physics Department, Yarmouk University, Irbid 21163, Jordan
[a*]s.mahmood@ju.edu.jo (corresponding author), [b]jaqsaw_89@yahoo.com,
[c]ASA8120941@fgs.ju.edu.jo, [d]ahmadmoh@yahoo.com, [e]ibrahimbsoul@yahoo.com,
[f]amufeed@yu.edu.jo, [g]q.muhaidat@yu.edu.jo

**Key Words**: Y-type Hexaferrite, Structural Properties, Magnetic Properties, Mössbauer Spectroscopy

**Abstract**. The present study is concerned with the fabrication and characterization of Me$_2$Y substituted hexaferrites, Ba$_2$Me$_2$Fe$_{12-x}$T$_x$O$_{22}$ (Me = Co$^{2+}$, Mg$^{2+}$, and Cr$^{2+}$, and T = Fe$^{3+}$, and Ga$^{3+}$). The samples were prepared by the conventional ball milling technique and sintering at 1200° C. The effect of the choices of Me and T ions on the structural and magnetic properties of the hexaferrites were investigated. XRD patterns, magnetic parameters, and Mössbauer spectra of the Co$_2$Y were consistent with a single phase Y-type hexaferrite. However, the CoCr-Y sample was found to be dominated by the Y-type hexaferrite, and M-type and BaCrO$_4$ minority phases were observed in the XRD pattern of the sample. The small increase in saturation magnetization from about 34 emu/g up to 37.5 emu/g was therefore attributed to the development of the M-type phase. On the other hand, XRD pattern of the Cr$_2$Y sample indicated the dominance of the M-type phase in this sample. The high coercivity (1445 Oe) of this sample is evidence of the transformation of the material from a typically soft magnetic material (Y-type) to a hard magnet (M-type). The Ga-substitution for Fe in Co$_2$Y did not affect the saturation magnetization significantly, but the coercivity was reduced. However, the sample Ba$_2$CoMgFe$_{11}$GaO$_{22}$ exhibited a significant reduction of the saturation magnetization down to a value 26.6 emu/g, which could be due to the attenuation of the super-exchange interactions induced by the Mg$^{2+}$ substitution.

**Contents of Paper**

# 1. Introduction

Magnetic materials are widely used as essential components for a large variety of industrial and technological applications including, and not limited to, automotive industry, consumer electronic, data storage and processing, and next generation microwave (MW) devices [1-8]. Ferrites with a hexagonal structure (hexaferrites) were discovered in 1950s [9], and extensive research work involving the fabrication and control of the physical properties of these ferrites demonstrated their feasibility of replacing other ferrites and magnetic materials in a large variety of applications [10-12]. Among these hexaferrites, perhaps M-type hexaferrites are some of the earliest realized and most widely used for permanent magnet [13-15] and high density magnetic recording media [16-22].

The advancement in telecommunication technology, and the increased threat of electromagnetic interference (EMI) effects on the operation of devices call for the need for efficient electromagnetic absorbers at the operating frequencies of the devices [23]. Accordingly, significant efforts were made in pursuit of hexaferrite-based absorbing materials [24-32]. In addition, the need for device miniaturization and minimization of power consumption promoted the search for self-biased materials to be used in passive microwave devices [1,7, 33-36], where hexaferrites are practically irreplaceable for this purpose nowadays.

Efficient control of magnetic field by an applied electric field is essential for the production of materials for the newly evolving spin-electronic technology. Hexaferrites had demonstrated multiferroic properties and potential for spintronic applications at room temperature [37-40]. Specifically, Y-type hexaferrites were found to exhibit interesting multiferroic properties, and a significant amount of research work, both theoretical and experimental, was devoted to the investigation of the electronic and magnetic structure in these materials [41-51].

## 2. Classification and Properties of Hexaferrites

Barium hexaferrites are usually synthesized by high temperature sintering of a combination of BaO, MeO (Me = divalent metal ion), and $Fe_2O_3$. The type of hexaferrite is obtained by appropriate choices of the stoichiometric ratios of these oxides. Accordingly, six main types of hexaferrites were realized, namely, M-type ($BaFe_{12}O_{19}$), Y-type ($Ba_2Me_2Fe_{12}O_{22}$), W-type ($BaMe_2Fe_{16}O_{27}$), X-type ($Ba_2Me_2Fe_{28}O_{46}$), Z-type ($Ba_3Me_2Fe_{24}O_{41}$) and U-type ($Ba_4Me_2Fe_{36}O_{60}$). These ferrites are well documented and their properties discussed in the literature [1, 9, 10, 52]. Efforts were made to optimize the phase purity and physical properties of hexaferrites by adopting several chemical and physical processes. These include, and are not limited to, sol-gel method [53-55], co-precipitation [56-58], hydrothermal [59, 60], wet chemical mixing [61, 62], and high-energy ball milling [63-69]. The structure of hexaferrites is determined by the stacking sequence of S, R, and T structural blocks. The S block with chemical formula $Me_2Fe_4O_8$ consists of two hexagonal close-packed oxygen layers, the R block ($BaFe_6O_{11}$) consists of three layers with one oxygen ion in the middle layer replaced by $Ba^{2+}$ ion substitutionally, and the T block ($Ba_2Fe_8O_{14}$) consists of four layers with one $Ba^{2+}$ ion substituting an oxygen ion in each of the two middle layers [9, 52]. The small metal ions occupy interstitial positions with tetrahedral, octahedral, and bi-pyramidal coordinations with oxygen cations. The magnetic ions at these sites form the spin-up and spin-down sublattices, and the imbalance between the net magnetic moments of these sublattices is responsible for the magnetization of the hexaferrites. These magnetic ions are also responsible for the magnetocrystalline anisotropy, which in turn has a crucial impact on the coercive fields in these ferrites.

The two basic hexaferrites are the M-type and Y-type, and the other hexaferrites are obtained by different combinations of these ferrites and the spinel block. Table 1 summarizes the various types of hexaferrites with the structural stacking in the unit cell [10].

**Table 1.** Types of hexagonal ferrites

| Hexaferrite type | Combination | Molecular formula | Structural stacking |
|---|---|---|---|
| M | M | $BaFe_{12}O_{19}$ | RSR*S* |
| Y | Y | $Ba_2Me_2Fe_{12}O_{22}$ | TSTSTS |
| W | M+S | $BaCo_2Fe_{16}O_{27}$ | RSSR*S*S* |
| Z | M+Y | $Ba_3Co_2Fe_{24}O_{41}$ | RSTSR*S*T*S* |
| X | 2M+S | $Ba_2Co_2Fe_{28}O_{46}$ | RSR*S*S* |
| U | 2M+Y | $Ba_4Co_2Fe_{36}O_{60}$ | RSR*S*T*S* |

**2.1. M-Type Hexaferrites:** The most widely investigated hexaferrite is the M-type, whose unit cell is composed of the RSR*S* structural stacking (the star indicates rotation by 180° around the *c*-axis), with typical lattice parameters of $a$ = 5.88 Å and $c$ = 23.2 Å. This hexaferrite is a uniaxial hard ferromagnet with easy *c*-axis, and has a saturation magnetization of about 100 emu/g at zero temperature, which decreases down to 72 emu/g at 20° C [9]. The physical properties were modified by the substitution of Ba ions and/or Fe ions by other metal ions. Specifically, Ba ions in these ferrites were substituted by Sr [8, 70, 71], Pb [72-74], and Ca [21, 75-77]. On the other hand, the effects of a large variety of substitutions for Fe ions on the structural and physical properties of hexaferrites were reported. Enhancement of the coercivity of M-type hexaferrite was achieved by substituting Fe ions by trivalent metal ions such as Al, Ga, or Cr [3, 78-81]. However, the coercivity was reduced by substitutions of Fe ions by a large variety of metal ion combinations [3, 19, 67-70, 82-87].

**2.2. Y-Type Hexaferrites:** The unit cell of the Y-type hexaferrites contains three formula units, and is composed of stacking the S and T structural blocks in the sequence STS′T′S″T″, where the primes indicate rotation by 120° around the *c*-axis. The lattice parameter *a* is similar to that of M-type, while $c$ = 43.56 Å [9, 88]. This kind of ferrite is a soft magnetic material with planar magnetic anisotropy [10, 89]. The six interstitial crystallographic sites available for the small metal ions in the Y-type hexaferrite lattice are: two tetrahedral sites ($6c_{IV}$, $6c_{IV}^*$), and four octahedral sites ($3a_{VI}$, $18h_{VI}$, $6c_{VI}$, $3b_{VI}$). The locations of these sites in the structural blocks, their multiplicities per formula unit, and spin orientations of magnetic ions in these sites are illustrated in Table 2.

**Table 2.** Crystallographic sites in Y-type hexaferrites lattice, number of ions per molecule, and spin orientation of magnetic ions.

| Sublattice label | Crystallographic site | Block | Number of ions | Spin |
|---|---|---|---|---|
| **$6c_{IV}$** | tetrahedral | S | 2 | ↓ |
| **$3a_{VI}$** | octahedral | S | 1 | ↑ |
| **$18h_{VI}$** | octahedral | S-T | 6 | ↑ |
| **$6c_{VI}$** | octahedral | T | 2 | ↓ |
| **$6c_{IV}^*$** | tetrahedral | T | 2 | ↓ |
| **$3b_{VI}$** | octahedral | T | 1 | ↑ |

Although Ba-based Y-type hexaferrites ($Me_2Y$) are the most common among this class of ferrites [90, 91], a variety of ferrites were synthesized with Ba ions substituted by other ions such as Sr [48, 54, 92, 93], Ca [94], and Pb [95]. Modifications of the properties of Y-type hexaferrites were achieved by appropriate choice of the Me ion. Specifically, examples of the influence of such substitutions on the magnetic properties of substituted Y-type hexaferrites are presented in Table 3. It is worth mentioning here that the coercivity is not well determined for a given stoichiometry, since in addition, it depends on the shape and size of the particles, and on the fabrication technique. Large variations in the saturation magnetization, however, could be due to the presence of impurity phases in some prepared samples. Further, modifications of the properties of Y-type hexaferrites by substitutions for $Fe^{3+}$ ions were also reported [12, 35, 49, 51, 94, 96-103]. However, research work devoted to the effects of such substitutions in Y-type hexaferrites is relatively limited.

**Table 3.** Magnetic properties of Y-type hexaferrites at room temperature (superscripted values were obtained at different temperatures).

| Ferrite | $M_s$ (emu/g) | $H_c$ (Oe) | $T_c$ (°C) | Reference |
|---|---|---|---|---|
| Ba$_2$Co$_2$-Y | 34 | - | 340 | [9] |
| | 39[a] | - | - | [9] |
| | 35.3[b] | - | - | [91] |
| | 27.8 | 106.3 | - | [104] |
| | 21.9 | 48.2 | - | [95] |
| Ba$_2$Ni$_2$-Y | 24 | - | 390 | [9] |
| | 25[a] | - | - | [9] |
| | 25.5 | 205 | 386.6 | [89] |
| | 45 | - | - | [105] |
| Sr$_2$Ni$_2$-Y | 40.34 | 555.44 | | [93] |
| Ba$_2$Zn$_2$-Y | 42 | - | 130 | [9] |
| | 72[a] | - | - | [9] |
| | 69.6[b] | - | - | [91] |
| Ba$_2$Mn$_2$-Y | 31 | - | 290 | [9] |
| | 42[a] | - | - | [9] |
| Ba$_2$Mg$_2$-Y | 23 | - | 280 | [9] |
| | 29[a] | - | - | [9] |
| | 22 | 100 | 277 | [48] |
| | 22.78 | 31.35 | - | [106] |
| | 30.47[c] | 545.87 | - | [106] |
| Ba$_2$Cu$_2$-Y | 28[a] | - | - | [9] |

[a] denotes at 0 K, [b] at 5 K, and [c] at 4.2 K

In this article, the structural and magnetic properties of a series of Y-type hexaferrites with different choices of the Me divalent ion, and with special substitutions of trivalent metal cations for $Fe^{3+}$ ions are addressed. X-ray diffraction (XRD) was used to investigate the phase purity and structural characteristics of the hexaferrite. Scanning electron microscopy (SEM) was employed to examine the particle shape and morphology, the homogeneity, and local stoichiometry of the samples. Mössbauer spectroscopy (MS) was used to investigate the hyperfine interactions of representative sample, where the hyperfine properties were discussed in light of the structural characteristics. Finally, the magnetic properties of the samples were investigated using the vibrating sample magnetometry (VSM).

## 3. Experimental Techniques

**3.1. Sample Preparation:** The powder precursors of Ba$_2$Me$_2$Fe$_{12-x}$T$_x$O$_{22}$ Y-type hexaferrites were prepared by high energy ball milling mixtures of stoichiometric ratios of analytical grade barium carbonate and metal oxides in a Pulvaressitte-7 ball mill equipped with zirconia bowls and balls. Wet milling was performed in an acetone medium with a powder-ball mass ratio of 1:14. The mechanical grinding was carried out for 16 hours at a rotational speed of 250 rpm. The resulting wet mixture was left to dry at room temperature, and then about 1 g of the powder was dry-pressed into a cylindrical pellet (1.5 mm in diameter) in a stainless steel die under the influence of 4-ton force. The discs were subsequently sintered at temperatures of 1200 °C for 2 h in air atmosphere.

**3.2. Sample Characterization:** Rietveld refinement of the XRD patterns of the samples was carried out to determine the phase purity and structural characteristics of the hexaferrite phases. For this purpose, FULLPROF software was employed. The patterns were collected in the angular range 20° ≤ 2θ ≤ 70° in steps of 0.01° using XRD 7000-Shimadzu diffractometer equipped with Cu-K$_\alpha$ tube. The sample for XRD measurement was finely ground powder.

The particle size distribution and particle morphologies, as well as the local chemical stoichiometry were examined by SEM imaging, and the energy dispersive spectroscopy (EDS) facility in an FEI-Inspect F50/FEG electron microscope.

The magnetic measurements were performed using a commercial vibrating sample magnetometer (VSM, MicroMag 3900, Princeton Measurements Corporation) which provides a maximum applied field of 10 kOe. Hysteresis loops at room temperature were obtained for all samples, from which the magnetic parameters were derived. Magnetization measurements versus temperature at a constant applied field of 100 Oe (thermomagnetic curves) were carried out on representative samples. The samples for magnetic measurements were needle-like pieces cut from the sintered disc.

Finally, the hyperfine interactions in representative samples were investigated using $^{57}$Fe Mössbauer spectroscopy (MS). This spectroscopy is based on the local Mössbauer effect [107], which was extensively used to investigate the local chemical environment, valence state, and site symmetry of Fe in alloys [108-112], intermetallic compounds [113-115], geological samples [116, 117], garnets [118-121], and various types of ferrites [122-125]. Mössbauer spectra for the samples were collected over 512 channels using a conventional constant acceleration Mössbauer spectrometer operating with $^{57}$Co/Cr source. Room temperature Mössbauer spectrum of a foil of pure metallic iron was used for velocity calibration, and isomer shifts [108] were measured with respect to the centroid of the calibration curve. The sample for MS was in the form of a thin powder layer in a disk-like Teflon holder.

# 4. Results and Discussion

**4.1. XRD Analysis:** The patterns of the samples Ba$_2$Co$_{2-x}$Cr$_x$Fe$_{12}$O$_{22}$ ($x$ = 0, 1, and 2) are shown in **Error! Reference source not found.**. The structural analysis indicated that the sample with $x$ = 0 (Co$_2$Y) consists of a major Y-type phase, consistent with the standard (JCPD: 00-044-0206), and a secondary M-type phase whose pattern is consistent with the standard (JCPD: 00-043-0002). The presence of a small amount of M-type in this sample is manifested by the appearance (2 0 5) and (2 0 6) reflections of the M-type, between which the (0 2 10) reflection of the Y-type phase lies (see Fig. 2). Rietveld refinement resulted in lattice constants $a$ = 5.8633 Å, $c$ = 43.522 Å. These parameters are in agreement with previously reported values [90, 91, 106, 126]. The fraction of M-type phase in this sample was found to be 10%. The unit cell of Co$_2$Y constructed from the refined cell parameters is shown in Fig. 3 (top view and side view). The figure shows the hexagonal symmetry of the lattice, with little distortions.

The sample with $x$ = 1 (CoCr-Y) consists of three phases, the Y-type phase (Y), the M-type phase (M), and BaCrO$_4$ (Cr) phase whose pattern was found to be consistent with the standard pattern (JCPD: 00-015-0376). The increase in the fraction of the M-type in this sample is obvious. The lattice constants for the Y-type in this sample ($a$ = 5.8610 Å, $c$ = 43.4939 Å) are slightly smaller than those of the previous sample. If this reduction is to be attributed to the substitution of Co$^{2+}$ ions by smaller Cr$^{2+}$ ions, then we conclude that low-spin Cr$^{2+}$ ions ($r$ = 0.73 Å) substituted high-spin Co$^{2+}$ ions ($r$ =

0.745 Å) at octahedral sites [127]. The BaCrO$_4$ reflections are clear in the angular range below 30° as shown in Fig. 4.

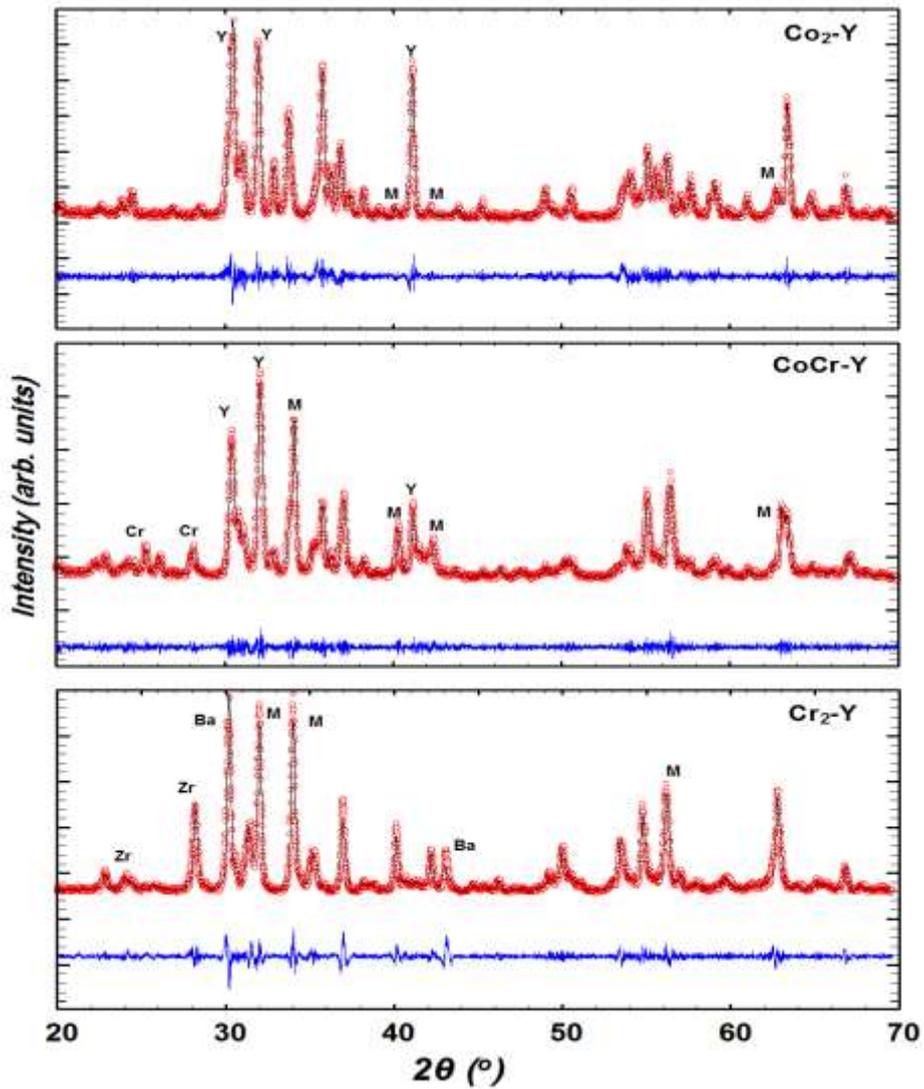

**Fig. 1:** XRD patterns with Rietveld refinement for Ba$_2$Co$_{2-x}$Cr$_x$Fe$_{12}$O$_{22}$ ($x$ = 0, 1, and 2)

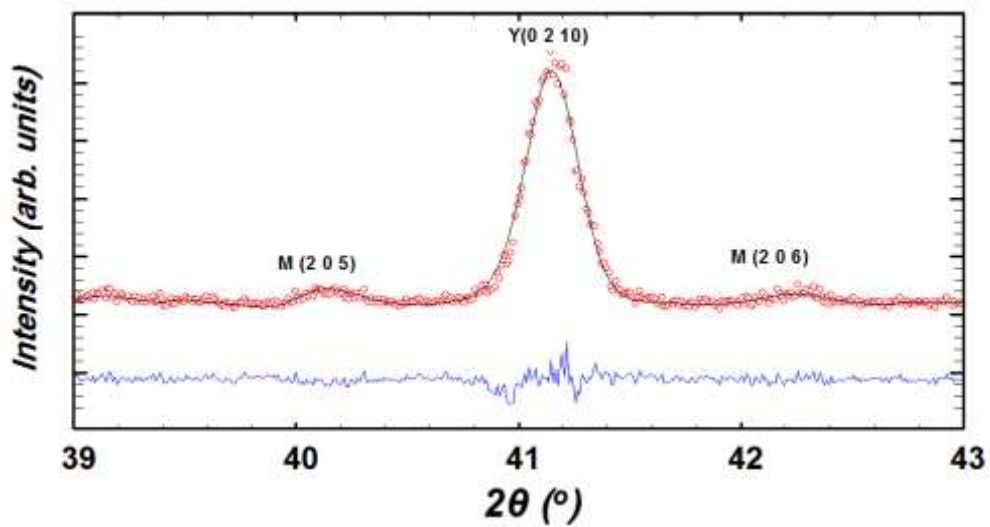

**Fig. 2:** Zoomed view of the XRD pattern with Rietveld refinement for Co$_2$Y sample showing evidence of a small amount of M-type.

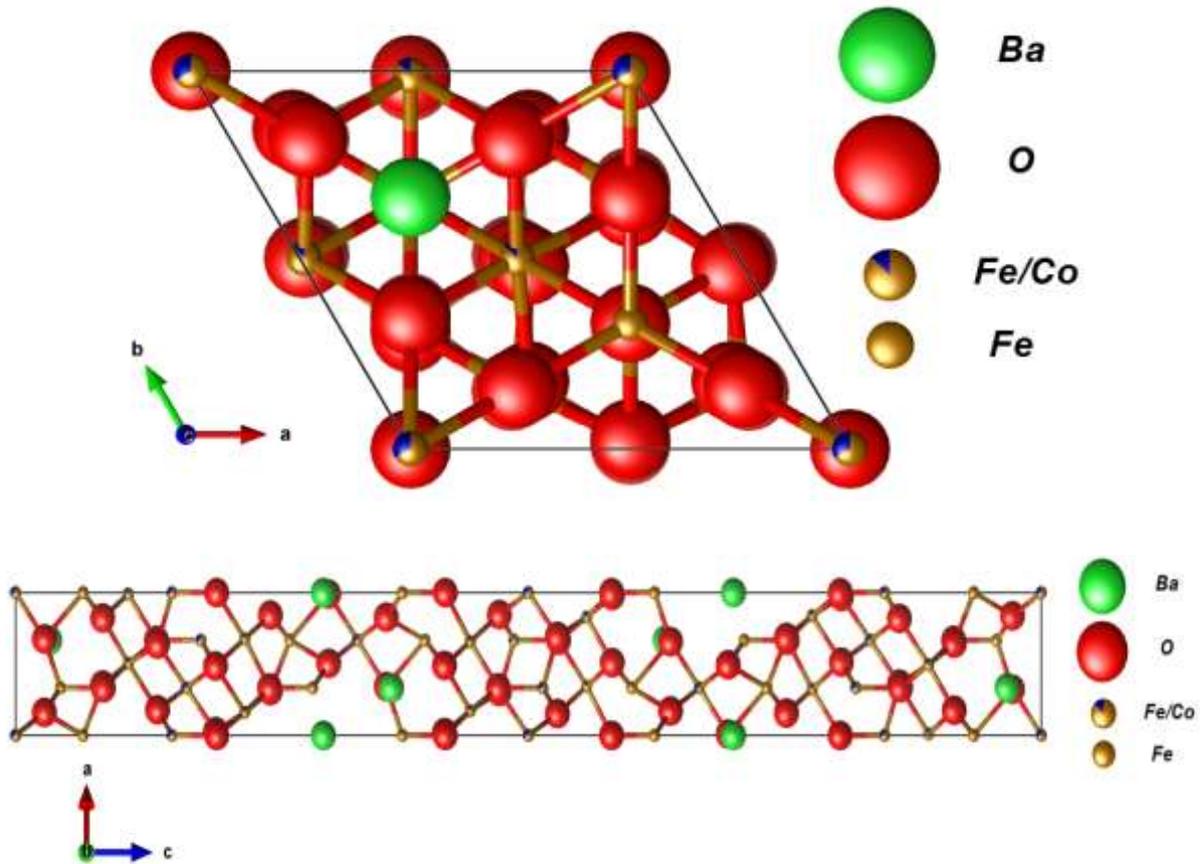

**Fig. 3:** Unit cell as viewed along the *c*-axis (upper panel) and along the *b*-axis (lower panel).

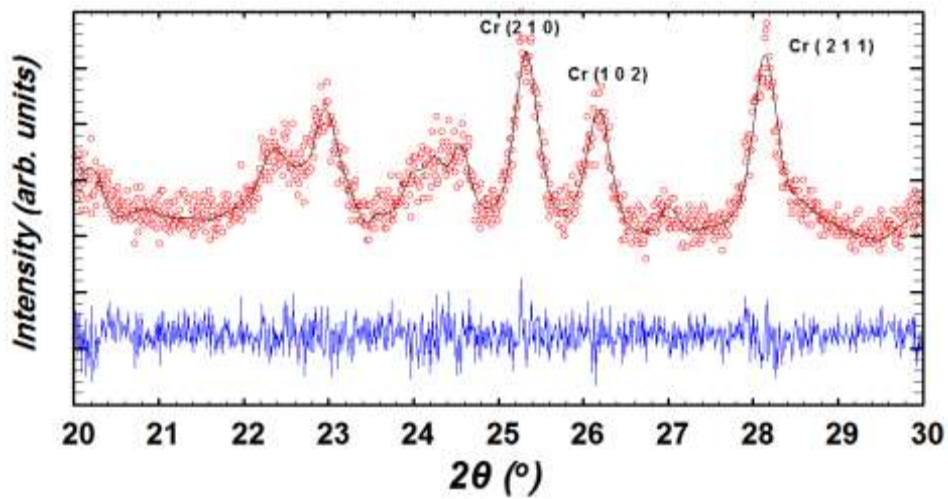

**Fig. 4:** Zoomed view of the diffraction pattern of CoCr-Y sample confirming the presence of the BaCrO$_4$ phase.

Further, the sample with $x = 2$, was found to be dominated by the M-type phase, with a strong evidence of the presence of barium-rich Ba$_3$Fe$_2$O$_6$ (Ba) phase, whose diffraction pattern was consistent with the standard pattern for this oxide (JCPD: 00-025-1477). Noting that the Ba/(Fe+Cr) molar ratio of 1:7 used in the starting powders is significantly higher than the molar ratio of 1:12 required for BaM, we conclude that the presence of a Ba-rich phase accompanying the BaM phase in this sample is natural. This result indicates that the equilibrium phases in this sample under the prevailing experimental conditions are distinctly different from those of Co$_2$Y or CoCr-Y samples. Also, strong evidence of $ZrO_2$ (Zr) phase was observed (Fig. 5), with a pattern consistent with the standard pattern (JCPD: 00-037-1484) for this oxide phase. The zirconium oxide phase in this sample may have resulted from contamination of the sample in the process of high temperature sintering in the zirconium oxide crucible.

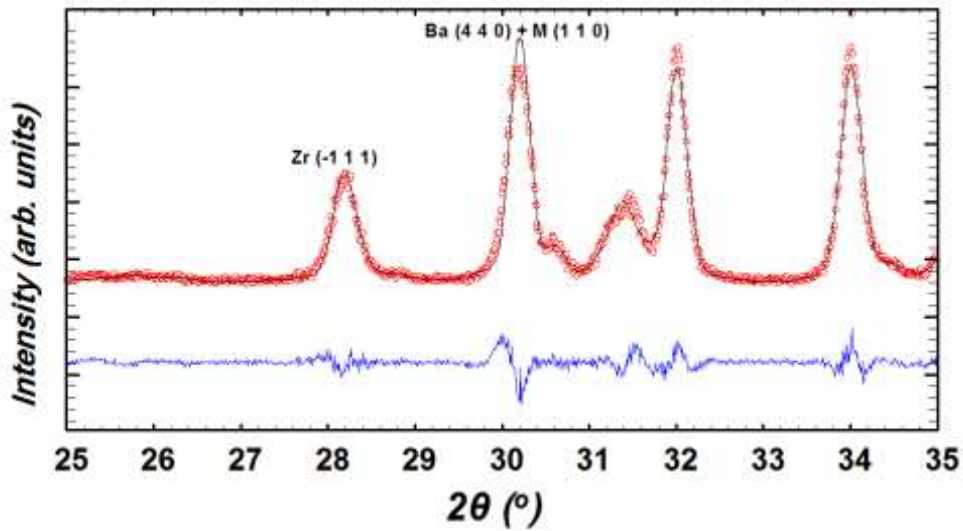

**Fig. 5:** Zoomed view of the diffraction pattern of $Cr_2Y$ sample confirming the presence of the $ZrO_2$ and $Ba_3Fe_2O_6$ phases.

The patterns of the samples $Ba_2Co_2Fe_{11}Ga_1O_{22}$ and $Ba_2CoMgFe_{11}GaO_{22}$ are shown in Fig. 6 and Fig. 7, respectively. Each of these samples was found to be composed of a major Y-type phase and a minor barium spinel ($BaFe_2O_4$) phase. The lattice parmeters for $Ba_2Co_2Fe_{11}GaO_{22}$ decreased slightly with respect to those of $Co_2Y$ sample. However, the lattice parameters for $Ba_2CoMgFe_{11}GaO_{22}$ increased (Table 4), demonstrating a noticeable rise in the *a* parameter as compared with that reported for $Mg_2Y$ [48, 106], while the value of *c* is consient with that reported by Koutzarova et al. [106], and higher than that reported by Khanduri et al. [94].

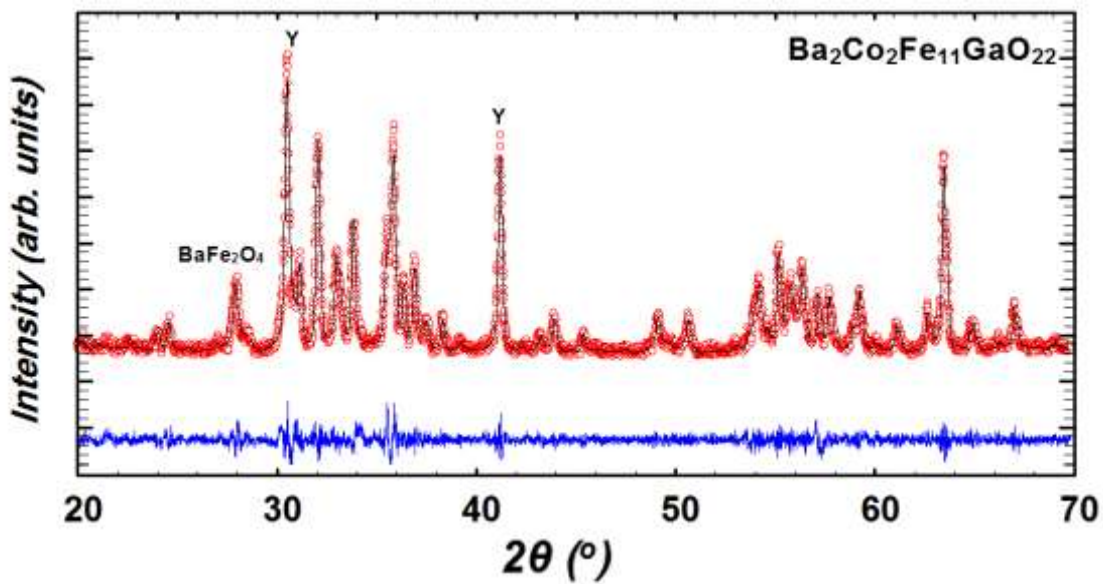

**Fig. 6:** XRD patterns with Rietveld refinement for $Ba_2Co_2Fe_{11}GaO_{22}$.

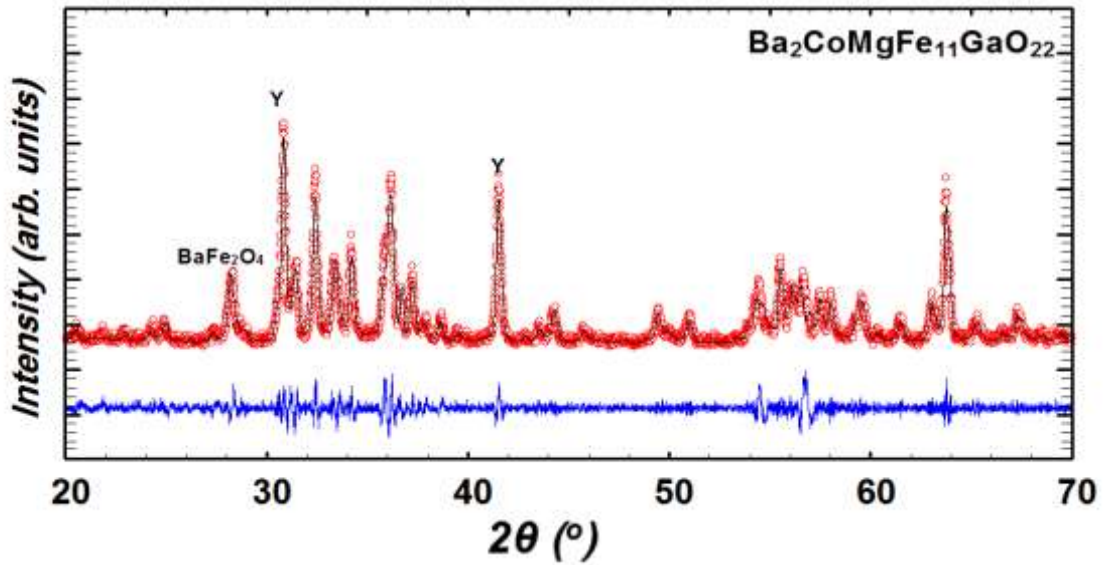

**Fig. 7:** XRD patterns with Rietveld refinement for Ba$_2$CoMgFe$_{11}$GaO$_{22}$.

**Table 4.** Structural information obtained by the XRD analysis of all samples.

| Sample | $\chi^2$ | $a$ (Å) | $c$ (Å) | $V$ (Å³) |
|---|---|---|---|---|
| Ba$_2$Co$_2$Fe$_{12}$O$_{22}$ | 1.29 | 5.8633 | 43.522 | 1295.8 |
| Ba$_2$CoCrFe$_{12}$O$_{22}$ | 1.10 | 5.8610 (Y) | 43.494 (Y) | 1293.9 (Y) |
| | | 5.8909 (M) | 23.243 (M) | 698.5 (M) |
| Ba$_2$Cr$_2$Fe$_{12}$O$_{22}$ | 1.85 | 5.9179 (M) | 23.357 (M) | 708.4 (M) |
| Ba$_2$Co$_2$Fe$_{11}$GaO$_{22}$ | 1.43 | 5.8585 | 43.486 | 1292.6 |
| Ba$_2$CoMgFe$_{11}$GaO$_{22}$ | 1.69 | 5.8906 | 43.4917 | 1306.9204 |

**4.2. SEM and EDS Analysis:** The Cr-substitution for Co in Co$_2$Y hexaferrite was found to induce significant structural changes in the different samples. These samples were consequently examined by SEM imaging to investigate the grain size and the microstructure of the samples. Further, Energy Dispersive Spectroscopy (EDS) measurements were carried out to investigate the variations of local stoichiometry in the samples, in an effort to confirm the presence of the equilibrium phases observed by XRD.

**4.2.1. Ba$_2$Co$_2$Fe$_{12}$O$_{22}$:** Representative images of the sample Ba$_2$Co$_2$Fe$_{12}$O$_{22}$ are shown in Fig. 8 below. The images indicated that the sample consisted of almost hexagonal platelets with a relatively wide distribution of particle size, and porous structure. The mean size of the plates was estimated to be in the range of 900-1700 nm. Much larger particles were also observed, with step-like formations.

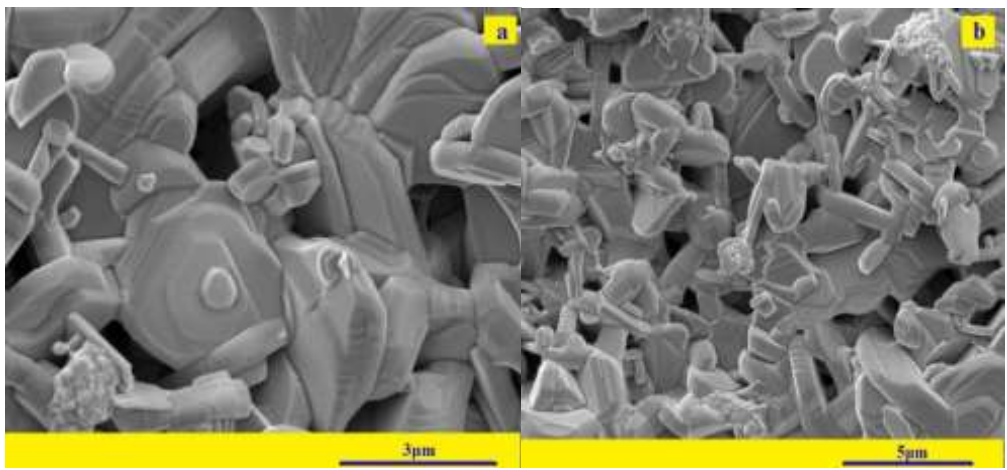

**Fig. 8:** SEM images for the sample $Ba_2Co_2Fe_{12}O_{22}$ at different scales

EDS measurements were made on different particles at different positions in the sample to examine the homogeneity of the sample. The results indicate that the atomic concentrations of Fe and Co (normalized to the atomic concentration of Ba) at one position are 6.17 and 0.98, respectively. These ratios are in good agreement with the stoichiometric ratios of 6 and 1 for $Co_2Y$ hexaferrite, which indicates incorporation of all ionic species in forming the Y-type phase at the measured position. However, measurements of the atomic ratios at a different position revealed lower concentration for Fe (5.45) and higher concentration for Co (1.63). The normalized atomic ratio of the two elements combined is 7.08, which is in very good agreement with the theoretical ratio of 7.00 for Y-type hexaferrite. This indicates that while the measured particle stoichiometry is evidence that the particle is composed of a single Y-type phase, the sample is not locally homogeneous.

**4.2.2. $Ba_2CoCrFe_{12}O_{22}$:** SEM images for this sample confirm the formation of well-crystallized hexagonal plates with a wide range of particle sizes (Fig. 9). The particles in this sample seem to have two distinct particle size distributions; one with mean particle diameter around 0.5 µm, and another with mean particle diameter clearly above 1 µm. The hexagonal plates, especially the large ones, are clearly thin, indicating in-plane preferential crystallization. Generally, the plates are stacked, with no noticeable porosity. Further, indication of the presence of impurity phases in the sample is revealed by the presence of regions with distinctly different grey levels (brighter and darker regions) as indicated by the images obtained in the EDS mode (Fig. 10).

EDS spectra at the darker and brighter regions of the sample are shown in Fig. 11, and the normalized atomic ratios are presented in Table 5. Fig. 11 shows that the relative intensity of the peak corresponding to Ba is significantly higher in the lighter regions than in the darker regions. The normalized atomic ratios of metals in the darker region is close to the stoichiometric ratios characteristic of Y-type, but generally lower, possibly due to the presence of a small amount of Ba-rich phase at the measured spot. The atomic ratios in the lighter region are, however, indicative of the presence Ba-rich phase in this region. The normalized atomic ratio of Cr (0.83) is in satisfactory agreement with the stoichiometry of $BaCrO_4$ phase detected by XRD. The Fe:Co atomic ratio (0.93:0.18) is 5.2:1, which is close to the stoichiometric ratio in $Co_2Y$ phase, and could indicate the presence of a small amount of Y-type phase at the measured spot. The Y-phase also contributes to the Ba spectral area corresponding to Ba atomic fraction similar to that of Co (0.18) in $Co_2Y$. Indeed, if this fraction is subtracted from the Ba concentration, the resulting Ba:Cr ratio in the Ba-rich phase would be 0.82:0.83, which is in excellent agreement with the stoichiometry of $BaCrO_4$ phase.

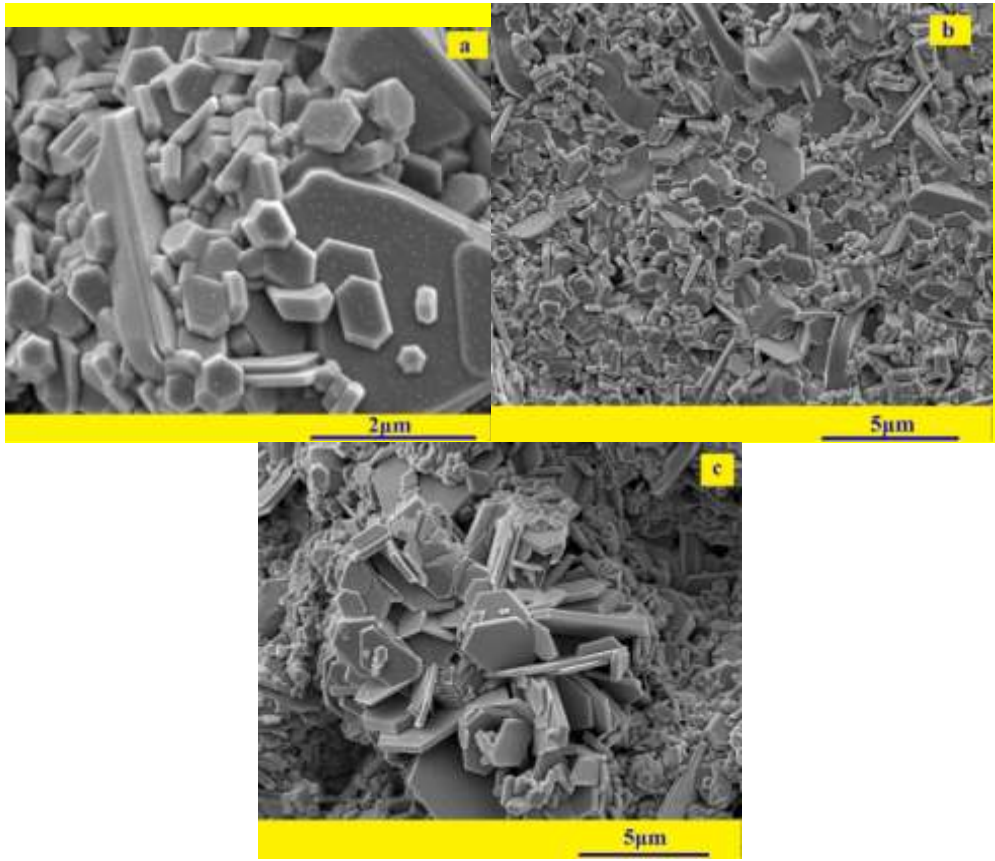

**Fig. 9:**1 SEM images for the sample Ba$_2$CoCrFe$_{12}$O$_{22}$ at different scales as indicated at the bottom of each image.

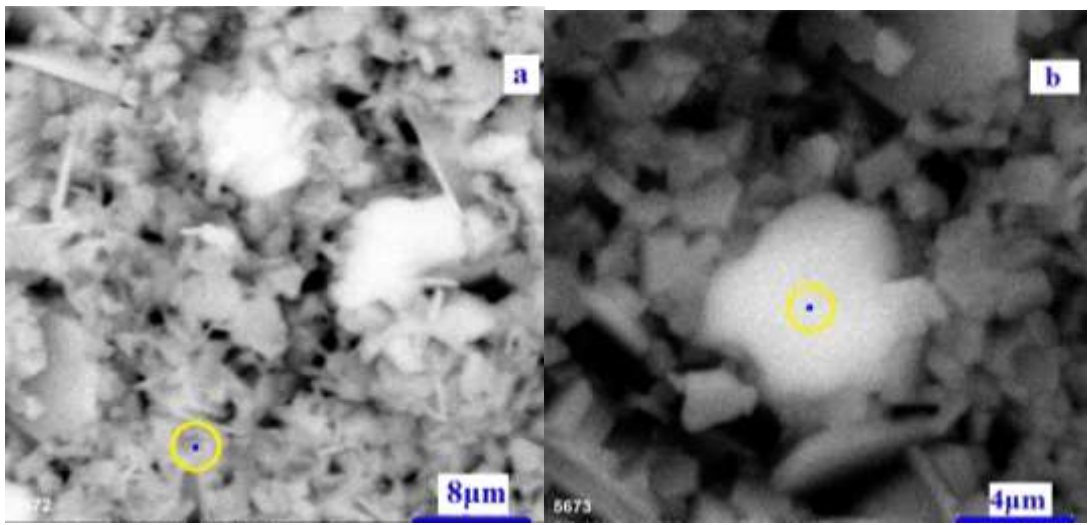

**Fig. 10:** SEM images of the sample Ba$_2$CoCrFe$_{12}$O$_{22}$ in the EDS mode.

**Table 5.** Normalized atomic ratios of metals in $Ba_2CoCrFe_{12}O_{22}$ at darker and lighter regions of the sample

| Position | Element | Normalized Atomic Ratio |
|---|---|---|
| Darker region | Cr | 0.63 |
| | Fe | 4.12 |
| | Co | 0.83 |
| | Ba | 1.00 |
| Lighter region | Cr | 0.83 |
| | Fe | 0.93 |
| | Co | 0.18 |
| | Ba | 1.00 |

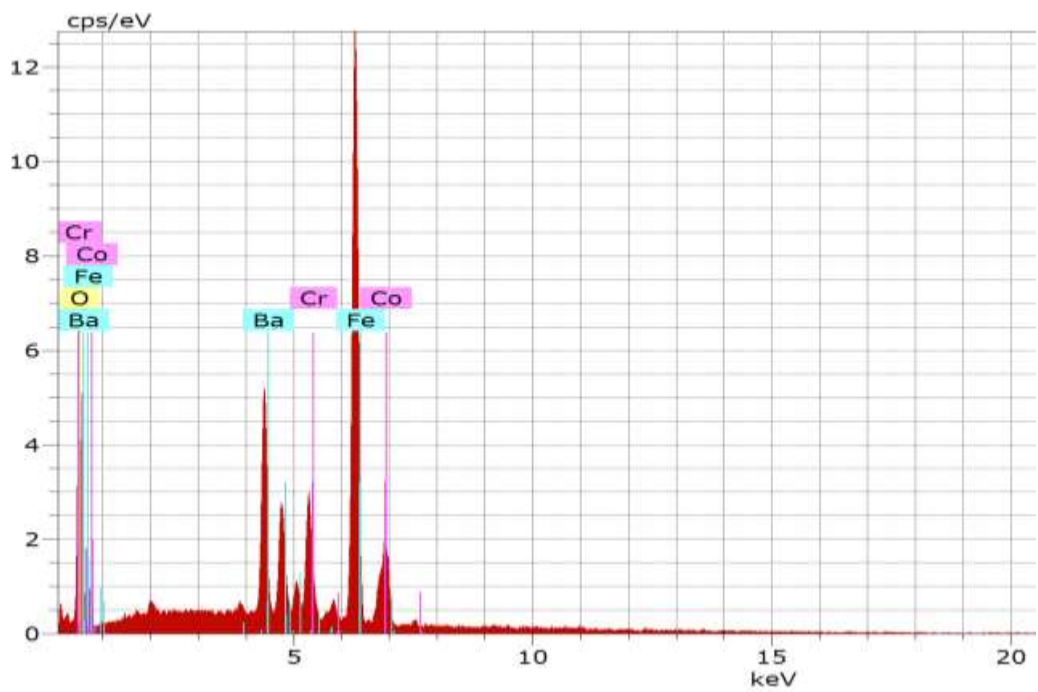

**(a)**

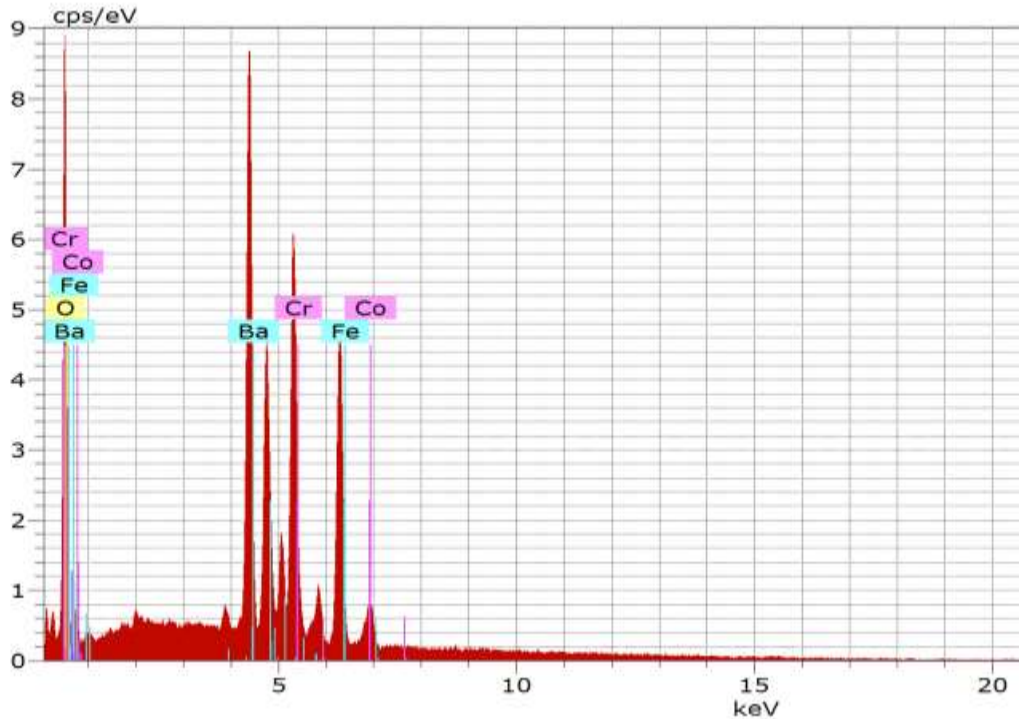

**(b)**

**Fig. 11:** EDS spectra of the sample Ba₂CoCrFe₁₂O₂₂ from **(a)** darker, and **(b)** lighter regions.

**4.2.3. Ba₂Cr₂Fe₁₂O₂₂**: Representative SEM images of the sample Ba₂Cr₂Fe₁₂O₂₂ are shown in Fig.12. Agglomeration of hexagonal-like particles with relatively narrow size distribution and mean diameter of about 0.5 µm were observed. In addition, smaller, non-hexagonal particles with diameters typically below 0.5 µm were observed. SEM images indicate the presence of different phases in this sample as evidenced by the different particle shapes, and the coexistence of darker and lighter particles in some regions of the sample. This argument is supported by the results of XRD measurements.

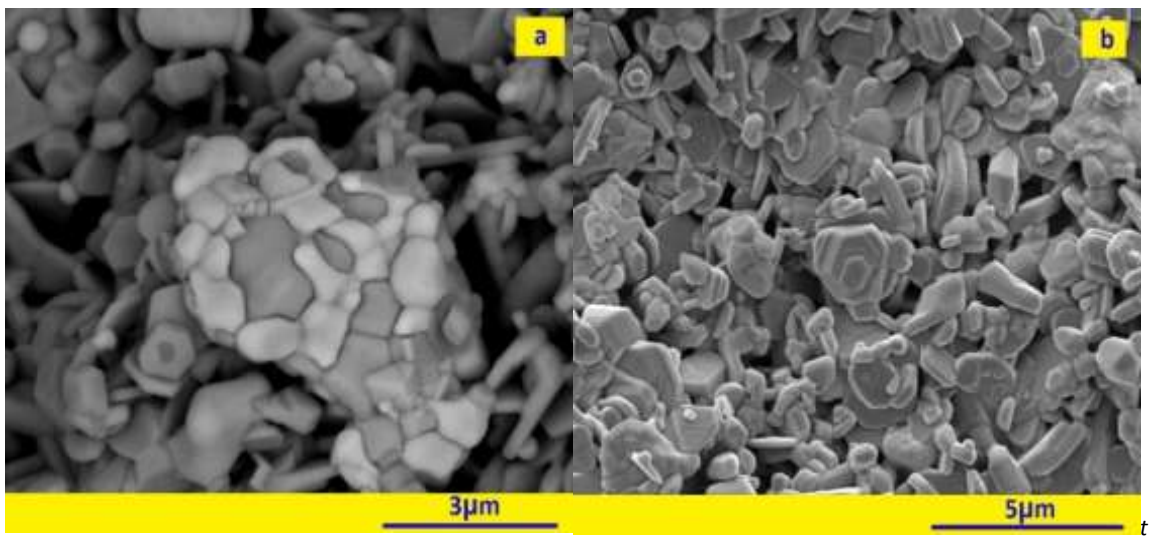

**Fig. 12:** Representative SEM images for the sample Ba₂Cr₂Fe₁₂O₂₂.

EDS measurment on a hexagonal plate revealed normalized atomic ratios of Fe and Cr of 11.61 and 1.18, respectively. These ratios are consistent with the stoichiometric ratios of BaM phase, and the EDS results, therefore, confirmed that the hexagonal plates are composed of BaM phase, in agreement with XRD results. EDS measurements at other positions in the sample revealed lower normalized metal fractions, which could be associated with the coexistence of BaM and the Ba-rich ($Ba_3Fe_2O_6$) phase (which was confirmed by XRD measurements) at the measured spot. EDS spectrum also showed clear evidence of Zr (Fig. 13), corresponding to the oxide phase detected by XRD.

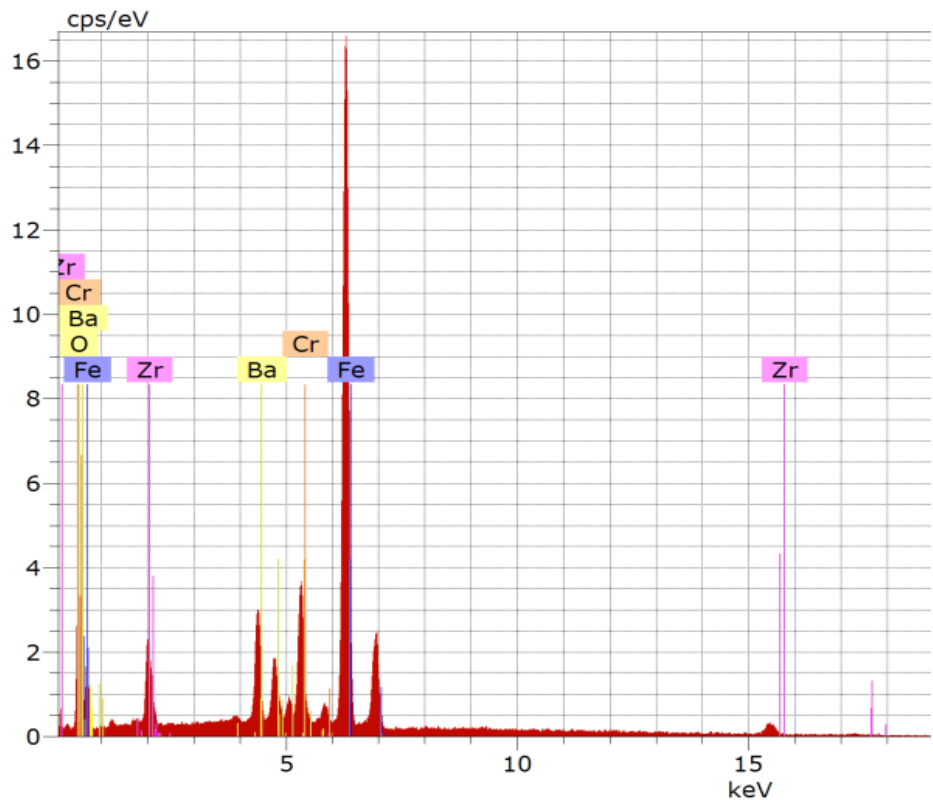

**Fig. 13:** EDS spectrum of the sample $Ba_2Cr_2Fe_{12}O_{22}$.

**4.3. Magnetic Measurements:** Magnetic hysteresis measurements were carried out on three samples of $Co_2Y$ hexaferrites with $Co^{2+}$ ions partially or fully substituted by $Cr^{2+}$ ions. Also, hysteresis measurements were carried out on a sample of $Co_2Y$ hexaferrite with $Fe^{3+}$ ions partially substituted by $Ga^{3+}$ ions ($Ba_2Co_2Fe_{11}GaO_{22}$) and on a sample with partial co-substitution of $Co^{2+}$ by $Mg^{2+}$, and $Fe^{3+}$ by $Ga^{3+}$ ($Ba_2CoMgFe_{11}GaO_{22}$). The saturation magnetization ($M_s$) was determined from the law of approach to saturation [67], while the coercivity ($H_c$) and remanent magnetization ($M_r$) were determined directly from the hysteresis loop. Further, thermomagnetic measurements were carried out on the Cr-substituted Y-type hexaferrites to investigate the magnetic phases and magnetic phase transition temperatures in these samples.

**4.3.1. Hysteresis Measurements:** The effect of the $Cr^{2+}$ ionic substitution for $Co^{2+}$ ions on the magnetic properties of $Co_2Y$ was investigated by carrying out magnetic hysteresis measurements on $Ba_2Co_2Fe_{12}O_{22}$, $Ba_2CoCrFe_{12}O_{22}$, and $Ba_2Cr_2Fe_{12}O_{22}$ samples (Fig. 14). The magnetic parameters derived from the hysteresis loops of the measured samples are shown in Table 6.

The saturation magnetization of $Co_2Y$ sample was found to be 33.9 emu/g, in very good agreement with the previously reported value of 34 emu/g [9]. However, this value is significantly higher than that reported by others [95, 104] (see Table 3). The coercivity of this sample is characteristic of a soft hexaferrite phase, and is in agreement with the value reported for $Co_2Y$ with similar particle size [104].

The sample $Ba_2CoCrFe_{12}O_{22}$ exhibited a small rise in saturation magnetization up to 37.5 emu/g, and a significant drop in coercivity down to 23 Oe. The increase in saturation magnetization could be partially due to the substitution of low-spin $Cr^{2+}$ ions for high-spin $Co^{2+}$ ions at spin-down sublattice, and partially due to the formation of M-type which has a higher characteristic saturation magnetization than Y-type hexaferrite. The significant drop in coercivity could be due to the attenuation of the magnetocrystalline anisotropy induced by the substitution of $Co^{2+}$ ions (which has a high contribution to magnetocrystalline anisotropy) by $Cr^{2+}$ ions. This reduction could also be associated with the differences in particle size and morphology as confirmed by SEM images.

The saturation magnetization of the $Ba_2Cr_2Fe_{12}O_{22}$ is also slightly higher than that of $Co_2Y$, and the relatively high coercivity (1445 Oe) is indicative of transformation to a harder magnetic phase in this sample. This result is consistent with the XRD patterns which confirmed the dominance of M-type in this sample. The coercivity in this sample, however, is significantly lower than that for a typical M-type hexaferrite (few thousand Oe). This could be associated with the attenuation of the superexchange interactions as a result of the Cr-substitution, which also leads to a reduction in saturation magnetization. The reduction in saturation magnetization down to 35.1 emu/g in this sample (with respect to 72 emu/g for BaM) is also associated with the presence of significant amounts of non-magnetic phases ($Ba_3Fe_2O_6$ and $ZrO_4$) in this sample.

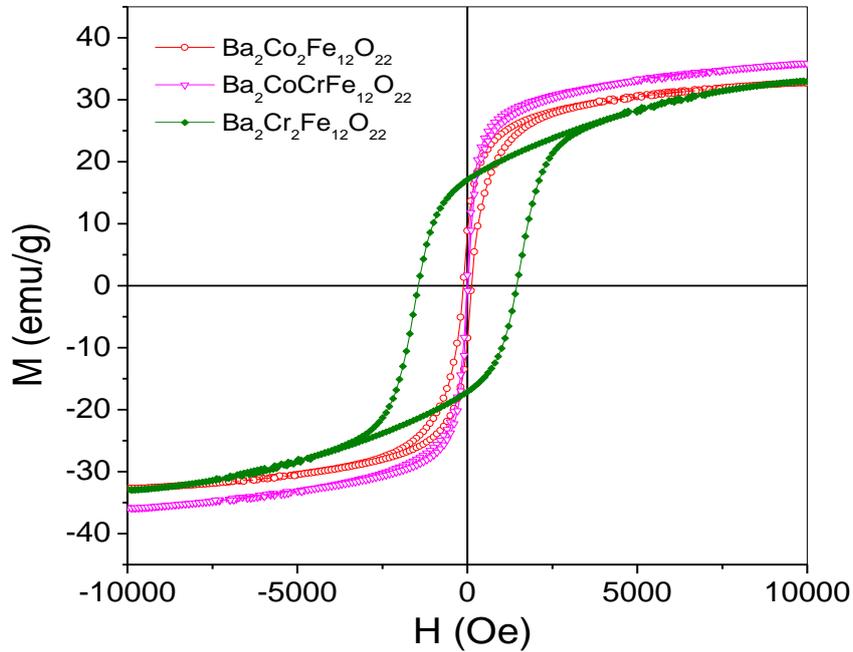

**Fig. 14:** 2 Hysteresis loops of Cr-substituted $Co_2Y$ samples.

**Table 6.** Magnetic parameters of the hexaferrites

| Sample | *Ms* (emu/g) | *Hc* (Oe) | *Mr* (emu/g) | *Mr/Ms* |
|---|---|---|---|---|
| $Ba_2Co_2Fe_{12}O_{22}$ | 33.9 | 123 | 7.90 | 0.23 |
| $Ba_2CoCrFe_{12}O_{22}$ | 37.5 | 23 | 1.58 | 0.04 |
| $Ba_2Cr_2Fe_{12}O_{22}$ | 35.1 | 1445 | 17.1 | 0.49 |
| $Ba_2Co_2Fe_{11}GaO_{22}$ | 34.9 | 88 | 6.88 | 0.20 |
| $Ba_2CoMgFe_{11}GaO_{22}$ | 26.6 | 81 | 5.81 | 0.22 |

The effect of Ga substitution for Fe, and Mg for Co on the magnetic properties of $Co_2Y$ hexaferrite is demonstrated by Fig. 15. The saturation magnetization (34.9 emu/g) of the sample $Ba_2Co_2Fe_{11}GaO_{22}$ is consistent with the value characteristic of $Co_2Y$ hexaferrite [9], which was confirmed by XRD analysis. The fact that a secondary non-magnetic phase ($BaFe_2O_4$) was confirmed by XRD (which should result in a reduction of the saturation magnetization) makes us believe that Ga has a preference of substituting Fe at spin-down sites, thus compensating for the anticipated reduction. Further, the coercivity was lowered down to 88 Oe, which could be associated with the reduction of the strength of superexchange interactions as a result of substituting Fe ions by non-magnetic Ga ions. However, this coercivity value remained within the range of values reported for $Co_2Y$ hexaferrite [104].

The effect of Mg-substitution for Co in the sample $Ba_2CoMgFe_{11}GaO_{22}$ was found to reduce the saturation magnetization significantly (down to 26.6 emu/g). This reduction is consistent with the reported gradual reduction of the saturation magnetization of $Sr-Ni_2Y$ with Mg-substitution for Ni (down to 32.84 emu/g at $x = 0.5$) [93]. The saturation magnetization of our sample is higher than that (~23 emu/g) reported for $Mg_2Y$ hexaferrite (see Table 3), and lower than that of $Sr-Ni_2Y$ with 0.5 Mg substitution for Co. This result supports the argument that the increase in Mg concentration results in a decrease in saturation magnetization. Further, the reduction of the coercivity (with respect to $Co_2Y$) down to 81 Oe could be associated with the reduction of the strength of the superexchange interactions in this sample [93].

With the exception of the $Cr_2Y$ sample, the squareness ratio (*Mr/Ms*) for all samples was rather low (below 0.25). This is indication of the multi-domain structure of the magnetic particles, where the magnetization processes are dominated by magnetic domain-wall motion rather than by magnetization rotation. This result is consistent with the large particle size observed in these samples in comparison with the critical single-domain size of less than 1 µm for hexaferrites [10]. The squareness ratio of the $Cr_2Y$ sample, however, is very close to the 0.5 value characteristic of an assembly of randomly orientated single-domain magnetic particles. This is evidence that the magnetic behavior of this sample is dominated by the small (~ 0.5 µm) BaM particles observed by SEM in this sample.

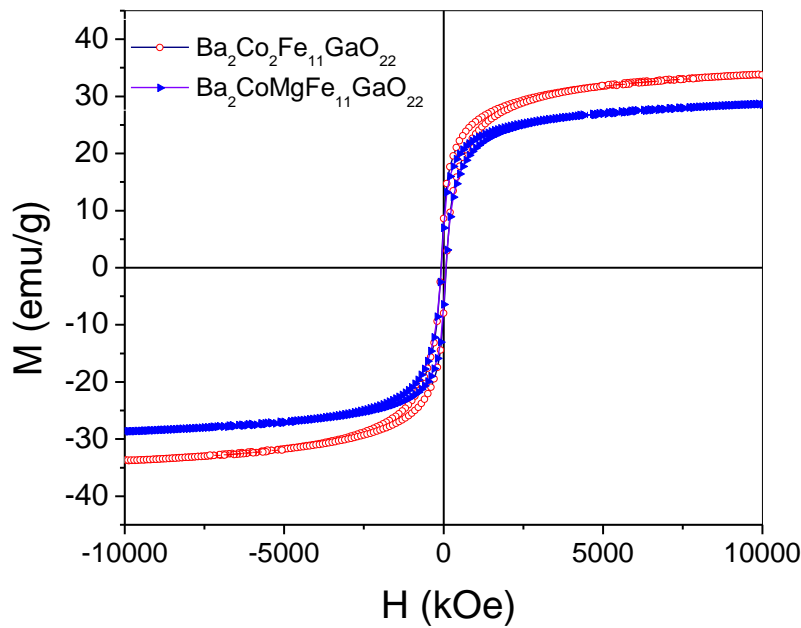

**Fig. 15:** Hysteresis loops of the samples $Ba_2Co_2Fe_{11}GaO_{22}$ and $Ba_2CoMgFe_{11}GaO_{22}$.

**4.3.2. Thermomagnetic Measurements of Cr-Substituted $Co_2Y$:** The critical transition temperature of a magnetic material can be determined from the temperature dependence of the magnetization at a fixed applied magnetic field. Consequently, thermomagnetic measurements provide information on the magnetic phases with different critical temperatures coexisting in the sample. To investigate the effect of Cr substitution on the phase evolution in $Co_2Y$ hexaferrites, thermomagnetic measurements under an applied field of 100 Oe were carried out on the $Ba_2Co_{1-x}Cr_xFe_{12}O_{22}$ ($x = 0, 1, 2$) samples.

Fig16 shows the temperature dependence of the magnetization for the sample $Ba_2Co_2Fe_{12}O_{22}$. The thermomagnetic curve for this sample shows a sharp ferromagnetic-paramagnetic phase transition manifested by the rapid drop of the magnetization between 300 and 400° C. The transition temperature was determined from the derivative of the thermomagnetic curve. The strong (negative) peak confirmed that the sample was dominated by a single magnetic phase, and the narrow peak profile indicates magnetic homogeneity of the sample. From this result, we conclude that the chemical inhomogeneity detected by EDS measurements did not have significant effect on the broadening of the transition temperature. The Curie temperature for this phase was determined from the peak position and found to be 342° C, which is consistent with the value of 340 °C reported for $Co_2Y$ hexaferrite [9]. This result is consistent with the XRD results which indicated the presence of a Y-type hexaferrite in this sample. The small peak at a temperature of 527° C represents magnetic phase transition at a temperature significantly higher than the transition temperature of 450° C for pure BaM. A similar high temperature anomaly was observed in Co-Ti doped M-type hexaferrite, which was associated with the strengthening of the superexchange interactions induced by the substitution of $Co^{2+}$ ions in the lattice [127]. The position of this peak is, however, not fixed, and depends on the Co concentration in the sample. The appearance of this peak in this sample could therefore be associated with a minor Co-substituted M-type phase which was confirmed by XRD measurements. Since M-type has a significantly higher saturation magnetization than $Co_2Y$ (more than double), the presence of 10% of this phase in the sample (from XRD data) should lead to a significant increase in saturation magnetization with respect to $Co_2Y$, contrary to the observed value which is almost identical to that for $Co_2Y$. However, the magnetization of the high temperature phase was reported to be small [127], and therefore can hardly affect the magnetic properties of the sample.

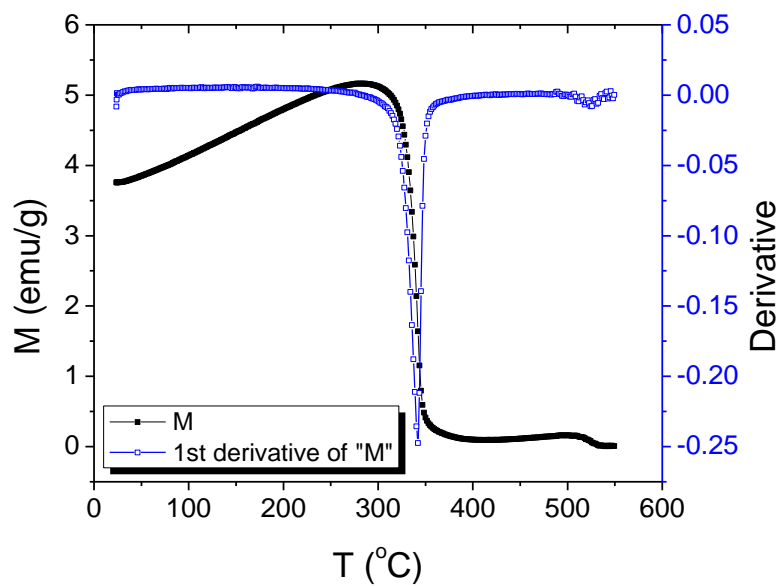

**Fig. 16:** Thermomagnetic curve measured at an applied field of 100 Oe, and its derivative for the sample $Ba_2Co_2Fe_{12}O_{22}$.

The thermomagnetic curve for the sample $Ba_2CoCrFe_{12}O_{22}$ clearly shows two transition temperatures corresponding to two different magnetic phases, in addition to the high-temperature anomaly at 554° C (Fig17). The positions of the two main dips in the derivative of the thermomagnetic curve are 339° C and 450° C, corresponding to $Co_2Y$ and M-type hexaferrite, respectively [9]. This result is consistent with the XRD results which confirmed the presence of M-type hexaferrite phase in this sample. The positive peak (Hopkinson peak) appearing in the magnetization curve near the Curie temperature of the M-type hexaferrite may indicate the presence of small particles of this phase which exhibit super-paramagnetic relaxations at this temperature [65]. The absence of such peak associated with the Y-type magnetic phase transition makes us believe that the two distinctly different particle distributions observed by SEM imaging correspond to the two structural and magnetic phases, with the small particles being M-type and the large particles being Y-type hexaferrites. The high temperature anomaly is again associated with the second magnetic phase transition in M-type resulting from the enhancement of the super-exchange interactions [127].

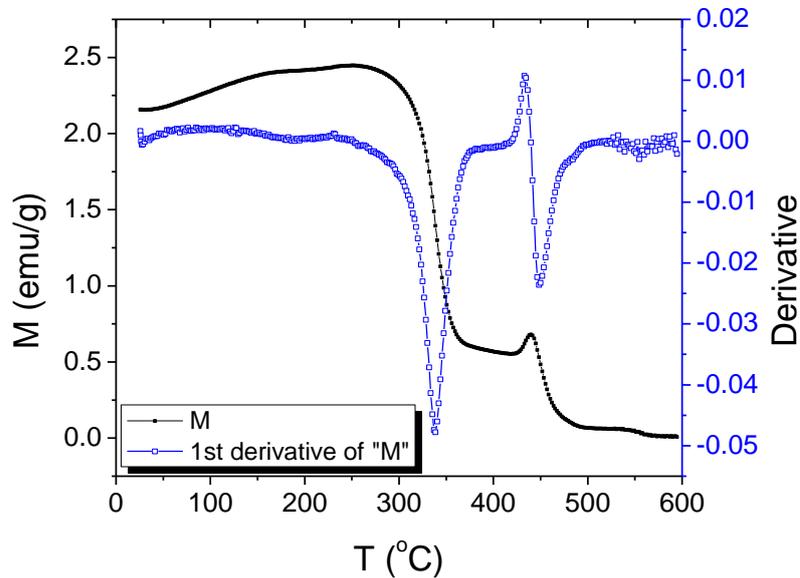

**Fig. 17:** Thermomagnetic curve and its derivative for the sample $Ba_2CoCrFe_{12}O_{22}$.

The thermomagnetic curve of the sample $Ba_2Cr_2Fe_{12}O_{22}$ is shown in Fig. 18. The figure clearly shows a Hopkinson peak near the critical transition temperature in the M-curve, which is associated with super-paramagnetic relaxations just below the critical temperature. The enhancement of the Hopkinson peak in this sample is consistent with the observed reduction in mean particle size as evidenced by SEM images. In addition, a small shoulder is evident at a temperature above 400° C. The derivative curve shows a negative peak at 339°, which is significantly less than that of pure BaM (450° C). This decrease in the transition temperature of the M-phase is correlated with the reduction of the superexchange interactions as a consequence of the Cr-substitution. This conclusion is supported by the observed lowering of the coercivity of this sample with respect to pure M-phase [66]. In addition, a small peak appeared at 425° C, which is within the range of the transition temperature of lightly-doped M-type phase. The presence of this peak could be evidence of phase separation in this sample, where islands of almost pure BaM phase are expected to be distributed in the main Cr-substituted BaM matrix. Further, the absence of the high temperature anomaly in this sample supports the previous conclusion that this anomaly is associated with the enhancement of superexchange interactions by Co ions.

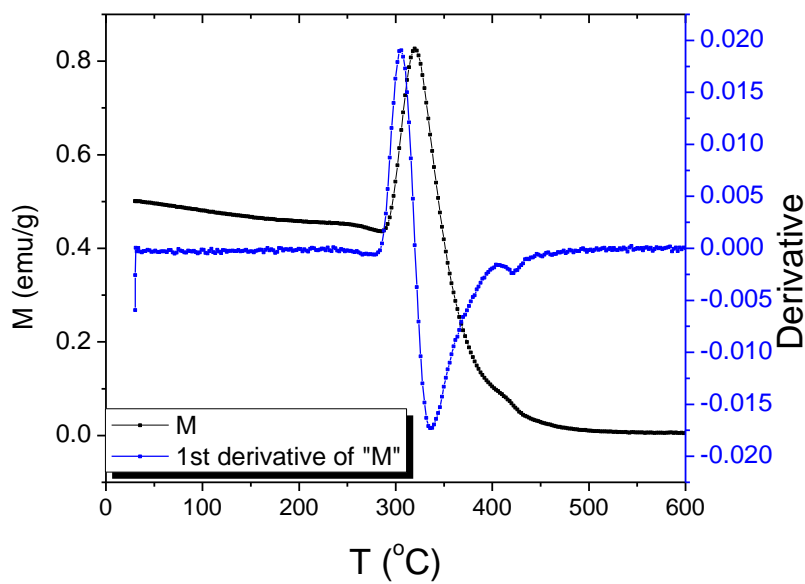

**Fig. 18:** Thermomagnetic curve and its derivative for the sample $Ba_2CoCrFe_{12}O_{22}$.

**4.4. Mössbauer Spectroscopy:** Although there are six different crystallographic sites which can be occupied Fe ions in the Y-type hexaferrite lattice, the spectral components corresponding to certain sites could not be resolved, resulting in a spectrum which can be satisfactorily analyzed in terms of only three components [54, 92, 124]. These components were variably assigned to different sets of crystallographic sites, but with a general agreement that the high-field component arises partially from Fe ions located at a tetrahedral site and partially from those located at the $3b_{VI}$ octahedral site. Also, there is an agreement between the different workers that the middle component is associated with the $18h_{VI}$ octahedral site. In a previous study [54], the high-field component in Mössbauer spectrum of $Co_2Y$ was argued to be associated with Fe ions at the $(6c^*_{IV} + 3b_{VI})$, while the low-field component was associated with $(6c_{IV} + 6c_{VI} + 3a_{VI})$ sites. In this study, $Co^{2+}$ ions were reported to occupy $6c_{VI}$ octahedral sites in the T block exclusively. The distribution of metal ions over the different sites is determined from the relative intensities of the sub-spectral components, which, in principle, are proportional to the number of Fe ions in the corresponding sets of sites [62].

**4.4.1. Ga-Substituted Y-type Hexaferrites:** Fig. 19 shows Mössbauer spectra of $Co_2Y$ with partial substitution of Co by Mg, and/or Fe by Ga. Generally speaking, the spectrum of the un-substituted sample is the same as that reported previously for $Co_2Y$ prepared by sol–gel method [54]. However, the substitution of Fe by Ga (in the sample $Ba_2Co_2Fe_{11}GaO_{22}$) resulted in a substantial reduction of the intensity of the high-field component whose outer absorption lines (indicated by the dashed drop lines) appeared as shoulders on the outer peak profile of the spectrum. This is evidence that at least partial substitution of Ga took place at sites corresponding to the high-field component, namely, $(6c^*_{IV} + 3b_{VI})$ [54]. However, a clear and more precise picture of the distribution of Ga ions in the sublattice can only be obtain by fitting the spectrum and obtaining the relative intensities corresponding to the different crystallographic sites.

The spectrum of the sample $Ba_2CoMgFe_{11}GaO_{22}$ also demonstrated reduction of the intensity of the high-field component, consistent with the substitution of Ga at $6c^*_{IV}$ site. Also, the hyperfine fields (indicated by the separations of the absorption lines) generally shifted to lower values with Mg-substitution. This reduction in hyperfine field is consistent with the reduction in magnetization for this sample, which is in agreement with the reported behavior of the hyperfine field with magnetization in ferromagnetic alloys [109].

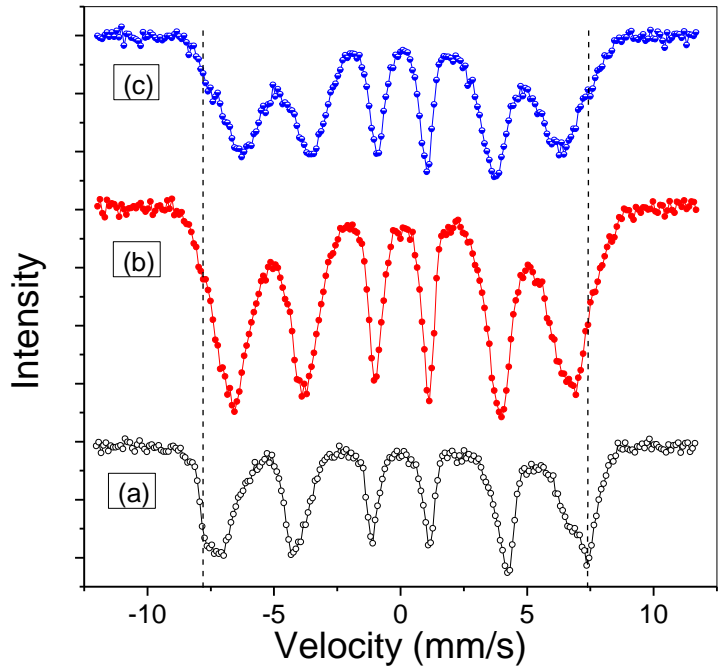

**Fig. 19:** Mössbauer spectra for (a) $Ba_2Co_2Fe_{12}O_{22}$, (b) $Ba_2Co_2Fe_{11}GaO_{22}$, and (c) $Ba_2CoMgFe_{11}GaO_{22}$.

The spectrum of $Ba_2Co_2Fe_{12}O_{22}$ sample was fitted with three magnetic components with zero quadrupole splitting (due to the symmetry of the absorption lines about the center of the spectrum, Fig. 20). The hyperfine parameters resulting from the fit are listed in **Error! Reference source not found.**Table 7. The three sub-spectral components with relative intensities of 25%, 50%, and 25% were assigned to the $(6c^*_{IV} + 3b_{VI})$, $18h_{VI}$, and $(6c_{VI} + 6c_{IV} + 3a_{VI})$ sites, respectively [54]. This indicates that 3 $Fe^{3+}$ ions per molecule occupy each of the $(6c^*_{IV} + 3b_{VI})$ and $(6c_{VI} + 6c_{IV} + 3a_{VI})$ groups of sites, whereas 6 $Fe^{3+}$ ions occupy the $18h_{VI}$ site. This is a clear indication that the $Co^{2+}$ ions occupy the sites corresponding to the component with the lowest hyperfine field. In light of the fact that ions with lower valence state should occupy face-sharing polyhedral (in order to reduce the electrostatic energy of the crystal), we could conclude that $Co^{2+}$ ions occupy the $6c_{VI}$ sites.

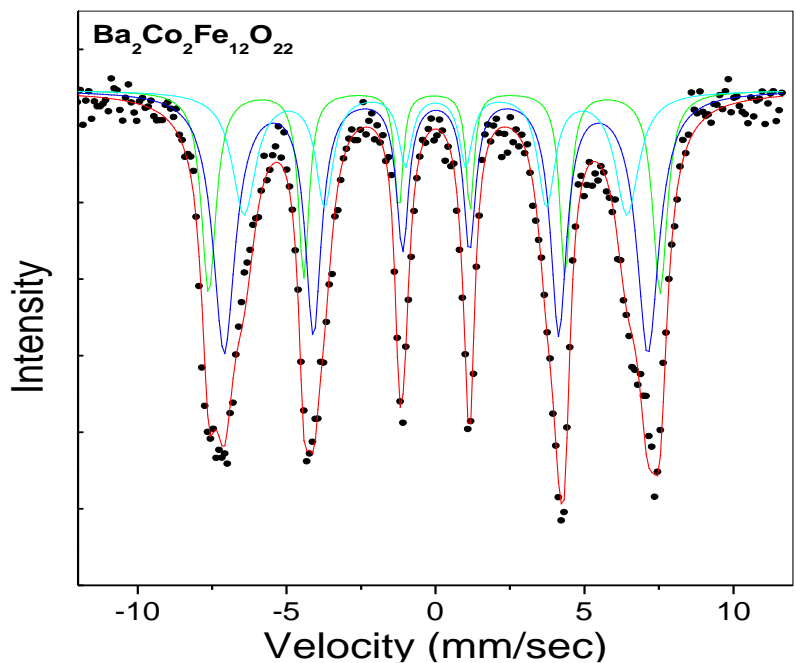

**Fig. 20:** Fitted Mössbauer spectrum of $Ba_2Co_2Fe_{12}O_{22}$

The spectrum of the sample $Ba_2Co_2Fe_{11}GaO_{22}$ was also fitted with three magnetic components with hyperfine parameters shown in Table 7. The relative intensities of 24%, 51%, and 25% of the ($6c_{IV}^*$ + $3b_{VI}$), $18h_{VI}$, and ($6c_{VI}$ + $6c_{IV}$ + $3a_{VI}$) components correspond to 2.6, 5.6, and 2.8 $Fe^{3+}$ ions per molecule in these sublattices, respectively. These numbers were obtained by the multiplication of the relative intensity of a given component by the total number of Fe ions (which is 11 in the sample). This result suggests that 0.4 Ga ions occupy the $6c_{IV}^*$ spin-down sublattice, and 0.4 Ga ions occupy the $18h_{VI}$ spin-up sublattice, a distribution which does not affect the net magnetic moment per molecule. Therefore, considering the results of the magnetic measurements on this sample, it is safe to conclude that the remaining 0.2 Ga ions per molecule in the third set of sites occupied the $6c_{IV}$ spin-down site, since the $6c_{VI}$ spin-down site is fully occupied by Co ions [54]. Thus, Ga ions in this sample are distributed at both spin-up and spin-down sites, with preference for tetrahedral spin-down sites in the S and T blocks of the Y-type hexaferrite lattice. The hyperfine fields of the three magnetic components clearly decreased with respect to the Co₂Y sample.

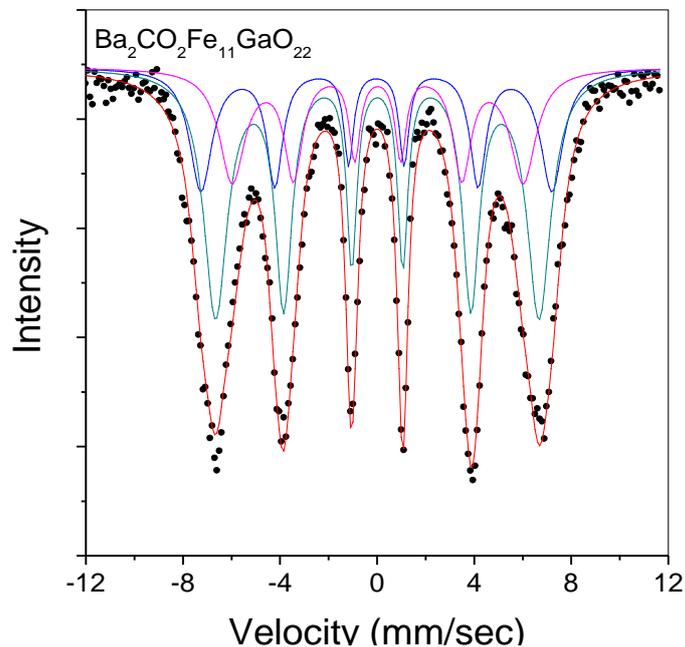

**Fig. 21:** Fitted Mössbauer spectrum of $Ba_2Co_2Fe_{11}GaO_{22}$

The fitted spectrum of the sample $Ba_2CoMgFe_{11}GaO_{22}$ is shown in Fig. 22. The hyperfine fields decreased further in this sample, reflecting a larger effect of the substitution of non-magnetic ions on the net magnetic field sensed by the Mössbauer probe (the $^{57}$Fe nucleus). The relative intensities of 19%, 55%, and 26% of the ($6c_{IV}^*$ + $3b_{VI}$), $18h_{VI}$, and ($6c_{VI}$ + $6c_{IV}$ + $3a_{VI}$) components correspond to 2.1, 6.0, and 2.9 $Fe^{3+}$ ions per molecule in these sublattices, respectively. This result indicates that the substitution of Ga does not occur at $18h_{VI}$ site in this sample. The significant decrease in the number of $Fe^{3+}$ ions in the ($6c_{IV}^*$ + $3b_{VI}$) sites indicates higher substitution of non-magnetic ions at these sites, displacing Fe ions to fill up the $18h_{VI}$ site. This substitution for Fe ions may occur partially at the $3b_{VI}$ site, which is responsible for the reduction in the strength of the superexchange interactions as argued in the context of discussing the magnetic data of this sample.

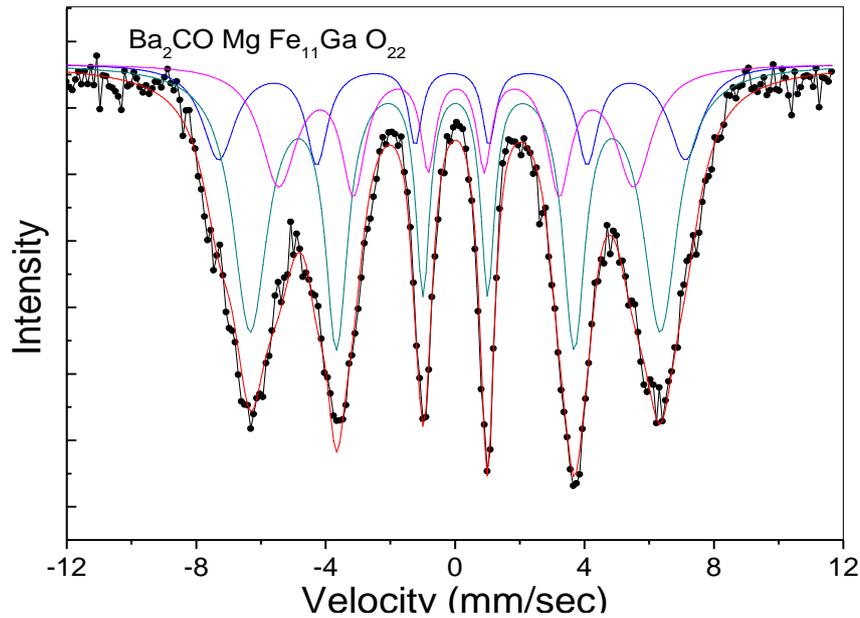

**Fig. 22:** Fitted Mössbauer spectrum of Ba$_2$CoMgFe$_{11}$GaO$_{22}$

**4.4.2. Cr-Substituted Y-type Hexaferrites:** Fig. 23 shows Mössbauer spectra of Co$_2$Y with partial or total substitution of Co by Cr. The figure clearly indicates that Cr substitution induces significant changes in the structural and magnetic phases as evidenced by the evolution of new spectral components. Specifically, the appearance of new components in the high-field region, and the sharp and strong absorption lines in the Cr$_2$Y sample (indicated by the dashed drop lines), are clear indications of the structural transformation in the samples. The sharp absorption peaks indicated by the dashed lines are characteristic of M-type phase [62, 71, 84]. The spectrum of the CoCr-Y sample, however, seems to be a combination of the spectra characteristic of M-type and Y-type hexaferrites, with modified hyperfine parameters due to Cr substitution. Since the hyperfine parameters of different magnetic phases with different structural symmetries are generally different, full analysis of the spectra of these samples was carried out in order to support the results of the structural and magnetic data.

The relatively complicated spectrum of the Cr$_2$-Y sample was fitted with five components (Fig. 24). The development of the first component with high hyperfine field (506 kOe), the strong component with hyperfine field of 418 kOe (characteristic of the 12$k$ sublattice of the M-type), and the weak component with high quadrupole splitting (associated with the 2$b$ bi-pyramidal site of the M-type) are clear indications of the structural transformation in this sample, which accords with the XRD and magnetic results. The main features of the hyperfine parameters of this sample are consistent with the parameters of M-type hexaferrite, with a noticeable reduction in the hyperfine fields of the outer components. A similar reduction was reported in TiRu-substituted SrM hexaferrite [71]. This reduction is consistent with the previously discussed reduction in saturation magnetization due to the weakening of the superexchange interactions in this sample.

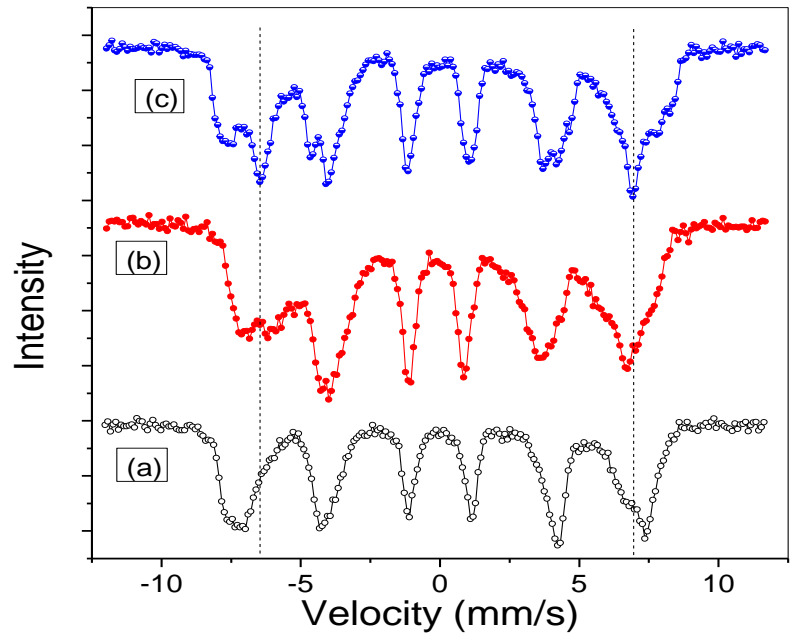

**Fig. 23:** Mössbauer spectra for (a) $Ba_2Co_2Fe_{12}O_{22}$, (b) $Ba_2CoCrFe_{12}O_{22}$, and (c) $Ba_2Cr_2Fe_{12}O_{22}$

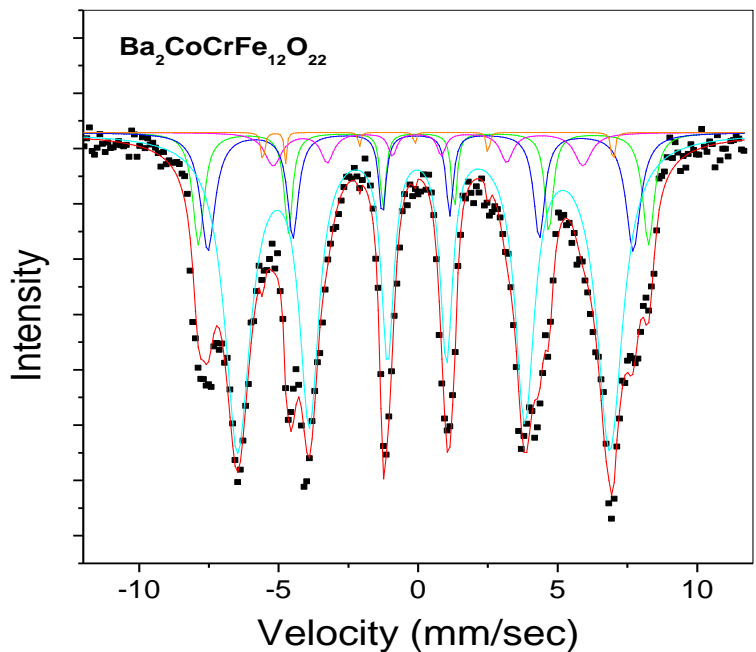

**Fig. 24:** Fitted Mössbauer spectrum of $Ba_2Cr_2Fe_{12}O_{22}$

The spectrum of the sample $Ba_2CoCrFe_{12}O_{22}$ was best fitted with four components (Fig. 25) with hyperfine parameters shown in Table 7. The general features of the hyperfine parameters suggest that the spectrum consists of a combination of the Y-type and M-type spectral components with lowered hyperfine fields (also see Fig. 23), which agrees with the structural refinement for this sample. The strong overlap between the spectral components, however, makes it difficult to resolve the individual sub-spectra, which may induce significant uncertainties in the derived hyperfine parameters. The main achievement of this analysis is therefore limited to the observation of the clear differences in the spectral characteristics of the samples, which correlate with the coexistence of different hexaferrite phases with modified magnetic properties in these samples.

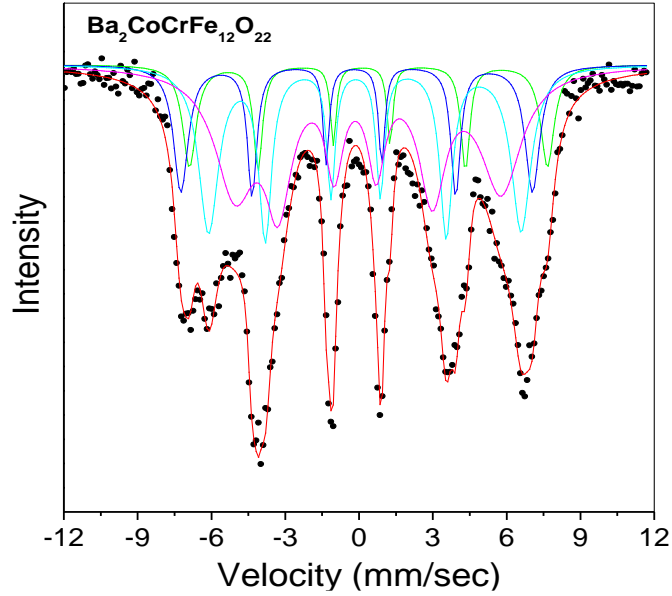

**Fig. 25:** Fitted Mössbauer spectrum of Ba$_2$CoCrFe$_{12}$O$_{22}$.

**Table 7.** Hyperfine parameters of substituted Co$_2$Y samples: hyperfine field ($B_{hf}$) isomer shift ($CS$), quadrupole splitting ($QQ$), and relative sub-spectral intensity ($A$) of the magnetic components.

| $B_{hf}$ (kOe) | $CS$ (mm/sec) | $QQ$ (mm/sec) | $A$ (%) |
|---|---|---|---|
| **Ba$_2$Co$_2$Fe$_{12}$O$_{22}$** | | | |
| 475 | 0.14 | - | 25 |
| 446 | 0.20 | - | 25 |
| 403 | 0.18 | - | 50 |
| **Ba$_2$Co$_2$Fe$_{11}$GaO$_{22}$** | | | |
| 543 | 0.15 | - | 24 |
| 418 | 0.18 | - | 51 |
| 376 | 0.20 | - | 25 |
| **Ba$_2$CoMgFe$_{11}$GaO$_{22}$** | | | |
| 452 | 0.08 | | 19 |
| 397 | 0.19 | | 55 |
| 344 | 0.22 | | 26 |
| **Ba$_2$Cr$_2$Fe$_{12}$O$_{22}$** | | | |
| 506 | 0.28 | 0.17 | 10.7 |
| 478 | 0.18 | 0.16 | 14.7 |
| 418 | 0.26 | 0.24 | 68.8 |
| 393 | 0.03 | 1.83 | 1.0 |
| 348 | 0.33 | 0.40 | 4.8 |
| **Ba$_2$CoCrFe$_{12}$O$_{22}$** | | | |
| 456 | 0.42 | 0.27 | 12 |
| 448 | 0.03 | 0.11 | 16 |
| 398 | 0.23 | 0.37 | 29 |
| 339 | 0.28 | 0.52 | 43 |

## 5. Conclusions

Structural refinements and magnetic measurements indicated that the substitution of Fe by Ga and/or Co by Mg in Co$_2$Y hexaferrites resulted in a major Y-type hexaferrite phase with slight improvement of the magnetic saturation due to the preferential substitution of Ga at spin-down sites. The substitution of Co by Mg, however, resulted in a significant drop in saturation magnetization which was attributed to the attenuation of the superexchange interactions between spin-up and spin-down sublattices. This reduction in saturation magnetization was accompanied by a noticeable reduction in the hyperfine fields of the Mg-substituted Y-type hexaferrite. The substitution of Co by Cr, however, induced major structural and magnetic changes in the samples. M-type hexaferrite phase, in addition to other nonmagnetic phases, evolved with the increase in Cr substitution. SEM, XRD, and thermomagnetic measurements on the sample with partial substitution of Co by Cr indicated phase segregation into M-type and Y-type, separately. The disappearance of the Y-type phase in the Co-fully substituted sample is an indication that the presence of Co is the major factor in the success in the production of a Y-type hexaferrite phase. In addition, the disappearance of the high-temperature peak in the derivative of the thermomagnetic curve of the Co-fully substituted sample is an indication that Co plays a major role in strengthening the superexchange interactions, resulting in a magnetic phase transition at elevated temperatures. Mössbauer spectrum of this (Cr2Y) sample was consistent with that of M-type hexaferrite, while the thermomagnetic measurements provided evidence of phase separation, indicating signatures of almost pure BaM islands in a heavily-Cr doped matrix.

## Acknowledgments

This work was supported by a generous grant from the Deanship of Scientific Research at the University of Jordan (Grant # 1404). The technical assistance of Y. Abu Salha and W. Fares (The University of Jordan) is acknowledged.